\documentclass[a4paper]{jpconf}
\usepackage{amsfonts, amsmath, amssymb, amsthm, amsbsy, amscd}

\newtheorem{prop}{Proposition}

\theoremstyle{definition}
\newtheorem{example}{Example}
\newtheorem{define}{Definition}

\theoremstyle{remark}

\newcommand{\pinner}{\mathbin{\mathchoice
   {\hbox{\vrule width0.6em depth0pt height0.4pt
   \vrule width0.4pt depth0pt height0.8ex}}
   {\hbox{\vrule width0.6em depth0pt height0.4pt
   \vrule width0.4pt depth0pt height0.8ex}}
   {\hbox{\kern0.14em
   \vrule width0.48em depth0pt height0.4pt
   \vrule width0.4pt depth0pt height0.6ex\kern0.14em}}
   {\hbox{\kern0.1em
   \vrule width0.39em depth0pt height0.4pt
   \vrule width0.4pt depth0pt height0.5ex\kern0.1em}}}}

\DeclareMathOperator{\ord}{ord}
\DeclareMathOperator{\img}{im}

\newcommand{\tu}{\tilde{u}}

\newcommand{\tv}{\tilde{v}}

\newcommand{\cE}{\mathcal{E}}

\newcommand{\veps}{\varepsilon}
\newcommand{\BBR}{\mathbb{R}}

\newcommand{\BBZ}{\mathbb{Z}}

\newcommand{\Id}{{\mathrm d}}

\newcommand{\jour}[1]{\textit{{#1}}}
\newcommand{\vol}[1]{\textbf{{#1}}}

\begin{document}

\title[Gardner's deformations as generators of new integrable systems]%
{Gardner's deformations as generators of new integrable systems}

\author{Arthemy V Kiselev$^1$ and Andrey O Krutov$^2$}

\address{$^1$ Johann Bernoulli Institute for Mathematics and Computer Science,
  University of Groningen,
  P.O.Box~407, 9700\,AK Groningen, The Netherlands.}
% \thanks{${}^{\dag}$%\address{
%   \textit{Current address}:
%   Institut des Hautes
%   \'Etudes Scientifiques,
%   Route de Chartres 35,
%   Bures-sur-Yvette, 91440 France%
% }

\address{$^2$ Department of Higher Mathematics, Ivanovo State Power
  University, Rabfa\-kov\-skaya str.~34, Ivanovo, 153003 Russia.
%  \textit{Tel.}:~+7(4932)269762
%  \textit{Fax.}:~+7(4932)326448
}

\ead{A.V.Kiselev@rug.nl, krutov@math.ispu.ru}

\begin{abstract}
We re-address the problem of construction of new infinite-dimensional completely integrable systems on the basis of known
ones, and we reveal a working mechanism for such transitions. By splitting the problem's solution in two steps, we explain
how the classical technique of Gardner's deformations facilitates --~in a regular way~-- making the first, nontrivial move,
in the course of which the drafts of new systems are created (often, of hydrodynamic type). The other step then amounts to
higher differential order extensions of symbols in the intermediate hierarchies (e.\,g., by using the techniques of
Dubrovin \emph{et al.} \cite{DubrLiuZhang2006,Dubr2006CMP} and Ferapontov~\emph{et~al.}~\cite{FerapMoro,FerapMoroNovikov}).
\end{abstract}

%\subsection*{Introduction} 
\noindent%
The aim of this short note is to further and illustrate a practical concept which was outlined earlier in \cite{TMF2007}
within this conference series. Namely, we revisit the problem of integrable deformation of a given infinite-dimensional
system; the seminal paper was \cite{Gardner1967}. Much work towards description of the arising moduli spaces has been 
performed by Dubrovin \emph{et al.}~\cite{Dubr96}, cf.~\cite{DubrLiuZhang2006,Dubr2006CMP}. (It must be recalled that
cohomological theories in this context and organization of the moduli spaces are sensitive to the choice of admissible
classes of differential functions --~e.\,g., polynomial, rational, or analytic~-- in which such structures are sought for.)
In the world of integrable systems there is a closely related aspect of integrability-preserving transition between
(solutions to) systems of PDEs (e.\,g., via B\"acklund transformations, see~\cite{JKIgonin2002}; a different approach was
developed in~\cite{BorisovZykov98}). Here one could employ the `heavy artillery' \cite{JKIgonin2002,JKKerstenIgonin} of 
jet-bundle techniques for deformation of the Cartan structure elements in coverings over PDEs by using the
Fr\"olicher\/--\/Nijenhuis bracket, see~\cite{NonRemGrad} for illustration.

Having its roots in topological QFT and yet possessing numerous applications elsewhere, a task of extending low-order
hierarchies with higher-order symbols remains a topic of particular interest in the field (\cite{Dubr96}, also
\cite{FerapMoro,FerapMoroNovikov}). For instance, such is the approach to hydrodynamic-type systems viewed as the weak
dispersion limits of larger, initially concealed models. Not limited to the above-mentioned class of evolution equations,
this concept suits well for PDE systems of order ${}\geqslant2$ whenever those are taken as drafts for the (re)construction
of larger models; certain restrictions could be imposed by hand at exactly this moment in order to narrow, e.\,g., the
classes of solutions of the draft systems, cf.~\cite{FerapMoro,FerapMoroNovikov}. At the same time, there co-exist many
schemes for extension of the symbol for a given system (e.\,g., one follows the perturbative approach of~\cite{Dubr96}
or applies the Lax-pair based techniques from~\cite{KuperIrish}).

From a broader perspective, there arise two natural questions: which systems are proclaimed `interesting', thus delimiting
the sets of start- and endpoints in the proliferation schemes, and where one could take those `interesting' systems from~---
or pick the drafts of new interesting PDEs. Leaving now aside the ever-growing supply from Physics or a straightforward idea
of ploughing the available lists of already known integrable systems, let us focus on a self-starting, regular algorithm
which exploits the classical ideas from geometry of differential equations (\cite{ClassSym%=JK+Vinogradov'1997,2003
,GDE12%=IHES/M/12/13
,Olver}).

Specifically, we take the existence of infinitely many integrals of motion as a selection rule for nonlinear evolutionary
systems; by default, we shall always assume that the collection of conversed quantities at hand is maximal, that is, it can
not be extended within a class of conservation laws with local densities (otherwise, a count of infinities could become
risky). As a rule, such systems tend to be bi-Hamiltonian at least in the case when the spatial dimension $n$ is equal to 1,
with $x^1\equiv x$, see~\cite{Magri} and also~\cite{ArturSergyeyevBiHam}. Let us note also that a requirement of existence
of conserved quantities is, generally speaking, stronger than a `symmetry integrability' assumption~\cite{MikhailovSokolovShabat}. (However in applications it is often convenient to weaken the former requirement in favour of
the latter; we shall profit from the use of both approaches, see Example~\ref{ExDeformKdV} and 
Proposition~\ref{PropDeformKBous} in what follows.)

As soon as we agree to study only those evolutionary systems which admit infinite towers of integrals of motion, it is
natural to first ex- and then inspect the existence of a (much better if polynomial) recurrence relation between the 
integrals' conserved densities~\cite{TMF2007,Gardner1967,KuperIrish,KiselevKarasu2006,JMP2010}. This yields a regular
procedure for consecutive calculation of the integrals of motion on the basis of all previously known data by starting from
the `seed' constants. Let us emphasize that such relations between the densities are much more valuable and informative
than ordinary recursion operators ${R\colon\varphi_i\mapsto\varphi_{i+1}}$ for symmetries or say,
${R^{\dagger}\colon\psi_i\mapsto\psi_{i+1}}$ for the `cosymmetries' of evolutionary PDE; in a sense, every algorithm which
explicitly produces the densities contains the built-in homotopy formula for reversion of the variational derivative that
takes densities to the respective generating functions $\psi$, cf.~\cite[\S~4.2]{GDE12} and~\cite{Olver}.

The classical notion of Gardner's deformation ${\mathfrak{m}_{\veps}\colon\cE_{\veps}\to\cE}$ for a completely integrable 
system~$\cE$ was designed for serving exactly this purpose~\cite{Gardner1967}; in the course of years, it has become the
parent structure for a plethora of concepts ranging from the Lax pair to formal $\tau$-function, etc. (for a more general
approach to the geometry of Gardner's deformations see~\cite{TMF2007,JMP2010,JMP2012}). Quite remarkably, this good old 
construction (see Definition~\ref{DefGardner} below) also answers the second question which we posed so far:
for a system~$\cE$ which undergoes the deformation, this procedure yields ``promising'' drafts $\cE'$ of ``interesting''
new systems $\cE''$. We sketched this line of reasoning in~\cite{TMF2007} and we now discuss it in more detail. Our
surprising conclusion is that the world of completely integrable systems could be much more `tense' and regularly
organized than it may first seem; for the adjacency relations $\cE\to\cE''$ spin a web across that set, with topology still
to be explored; to the best of our knowledge, a study of the physical sense for a property of two models $\cE$ and $\cE''$ 
to be adjacent has not yet begun. 

\centerline{\rule{1in}{0.7pt}}

%\subsection*{Preliminares}
\noindent%
Let us first introduce some notation and conventions.
In what follows we consider systems of evolutionary partial differential equations
with two independent variables $x$ and~$t$ and $m$-tuples of unknowns
$u={}^t(u^1,\dots,u^m)$. %(or $u = {}^t(u,v)$ for $m=2$).
We denote the corresponding dependent variables in extended equation in
Gardner's deformation as $\tu$. We use capital letters for unknowns in
extensions of adjoint system.
%{Notation for dependent in extended eqn; CAPITAL letters for extension of adjoint}

\begin{define}[Classical Gardner's deformation~\cite{TMF2007,Gardner1967}]\label{DefGardner}
Let $\cE = \left\{ u_t = f(x, u, u_x, u_{xx}, \dots, u_k) \right\}$
be a system of evolution equations (in particular, a completely integrable system).
Suppose $\cE_\veps = \{ \tu_t = f_\veps(x, \tu, \tu_x, \dots,
\tu_{\tilde{k}}; \veps)  \mid f_\veps \in \img \tfrac{\Id}{\Id x} \}$
is a deformation of $\cE$ such that at each point $\veps \in
\mathcal{I}$ of an interval $\mathcal{I}\subseteq  \BBR$ 
there is the \emph{Miura contraction}
$\mathfrak{m}_\veps = \{ u = u(\tu, \tu_x, \dots, \veps)\} \colon
\cE_\veps \to \cE$. 
Then the pair $(\cE_\veps, \mathfrak{m}_\veps)$
is the (\emph{classical}) \emph{Gardner deformation} for system~$\cE$.
\end{define}

\begin{example}[\cite{Gardner1967}]\label{ExDeformKdV}
Consider the Koreweg\/--\/de Vries equation
\begin{equation}\label{kdv}
  \cE_{\text{KdV}_3} = \{ u_{t_3} = -u_{xxx} - 6uu_x \}.
\end{equation}
Its deformation found by Gardner is as follows: the extended equation
\begin{subequations}\label{kdvdef}
\begin{align}
\cE_{\text{KdV}_3;\veps} = & \left\{\tu_{t_3}  = - (\tu_{xx} + 3\tu^2 -
  2\veps^2\tu^3)_x \right\}, \label{kdvext}
\\
\intertext{is mapped to the initial KdV at $\veps=0$ by the Miura contraction}
\mathfrak{m}_{\veps} = & \left\{ u  = \tu - \veps \tu_x - \veps^2\tu^2
  \right\}\colon\cE_{\text{KdV}_3;\veps}\to\cE_{\text{KdV}}.\label{kdvmiura}
\end{align}
\end{subequations}
By expanding the left-{} and right\/-\/hand sides of ~\eqref{kdvmiura} in~$\veps$,
one obtains a well\/-\/known recurrence relation for densities~$\omega_k$ of 
the integrals of motion $\int_{-\infty}^{+\infty}\omega_k\,\mathrm{d}x$ for~\eqref{kdv},
\[%begin{align*}
\omega_0 = u,  \qquad \omega_1 = u_x,\qquad
\omega_k = \tfrac{\Id}{\Id x}\omega_{k-1}  + \sum_{i+j=k-2}\omega_i\omega_j,\qquad k\geqslant 2.
\]%end{align*}
% Let us finally recall that KdV equation~\eqref{kdv} and its extension~\eqref{kdvext} 
% are homogeneous with respect to the scaling weights $[u] =[\tu] = 2$,
% $[\Id/\Id x] = 1$, $[\Id/\Id t = 3]$, and $[\veps] = -1$.\marginpar{What for?}
\end{example}

We say that the coefficient $\varphi%_{k}
(\tu,\tu_x,\dots)$ of the highest power of~$\veps$ in the right\/-\/hand side of a 
\emph{polynomial} (in~$\veps$) Gardner's extension~$\cE_\veps$ determines the
\emph{adjoint} system $\cE' %_{k} 
= \{ \tu_{t_k}  = \varphi %_k 
\}$.

\begin{example}
The evolution equation 
\begin{equation}\label{adjkdv}
\tu_{t_3} = -6\tu^2\tu_x
\end{equation}
is adjoint to the Korteweg\/--\/de Vries equation~\eqref{kdv}.
\end{example}

We notice that the adjoint systems are often dispersionless, although this is not
always the case (cf.\ \cite{TMF2007} and~\cite{KiselevKarasu2006}). Let us now address 
the natural problem of extension of the new, adjoint equation --\,and its hierarchy
which appears by construction of Gardner's deformation\,-- by adding terms with
higher\/-\/order derivatives (in particular, by switching on the dispersion in~$\cE'$).
There are many techniques for solving this problem: a straightforward computational
algorithm, which does not require that the adjoint system be hydrodynamic type, is
illustrated in what follows by using the `symmetry integrability' 
approach~\eqref{adjkdv} and software~\cite{SsTools}; we then report on the second iteration
of such proliferation scheme and discuss the third and other steps to follow.

\begin{example}
It is readily seen that equation~\eqref{adjkdv} is the second element in the
infinite hierarchy of adjoint systems (corresponding to Gardner's extensions of
higher KdV flows, with the Miura contraction~$\mathfrak{m}_\veps$ common for all of them),
which is
\begin{align*}
{}&\vdots & {} &\vdots {} \\
\cE^\prime_{\text{KdV}_2} &= \{\tu_{t_5} = \tu^4\tu_x \}, &
  \varphi_5 &=  \tu^4\tu_x, \\
\cE^\prime_{\text{KdV}_1} &= \{ \tu_{t_3} = -6 \tu^2\tu_x \}, &
  \varphi_3 &=  -6\tu^2\tu_x, \\
\cE^\prime_{\text{KdV}_0} &= \{ \tu_{t_1} =  \tu_x \}, &
  \varphi_1 &= \tu_x. 
\end{align*}
Clearly, the scaling weights are not uniquely determined for the dependent variables 
in the adjoint hierarchy; for definition, set
$[U] = 1$, $[\Id/\Id x] = 1$, and $[\Id/\Id t_k ] =  (k-1)[U]+[\Id/\Id x] = k$,
which is consistent with the dynamics.
For all $k\in\mathbb{N}$, let us now list all scaling\/-\/homogeneous 
differential polynomials $f_k$ of weights $[f_k] = [U] + [\Id/\Id t_k] =
k + 1$ with undetermined coefficients, excluding at once those terms which are already
contained in the respective right\/-\/hand sides of the adjoint 
hierarchy~$\cE^\prime_{\text{KdV}_k}$. For instance, we let
\begin{align*}
&\vdots\\
f_5 &= q_5 U^6 + q_6 U^3U_{xx} + q_7 U^2U_{xxx}  + q_8 U^2(U_x)^2 +
q_9 UU_{4x} + q_{10}UU_{xx}U_x + q_{11}U_{5x} + q_{12}U_{xxx}U_x\\
{}&{}\quad + q_{13}(U_{xx})^2 + q_{14}(U_x)^3 ,\\
f_3 &= q_1 U^4 + q_2 U U_{xx} + q_3 U_{xxx} + q_4 (U_x)^2,\\
f_1 &= 0.
\end{align*}
The Ansatz for the full hierarchy is thus
\begin{align*}
U_{t_k} = {} & \varphi_k + f_k %= \dots
, \\
\dots&{}\\
U_{t_5} = {} & \varphi_5 + f_5 = U^4U_x + q_5 U^6 + q_6 U^3U_{xx} + q_7 U^2U_{xxx}  + q_8 U^2(U_x)^2 +
q_9 UU_{4x} + q_{10}UU_{xx}U_x \\
{}&{} + q_{11}U_{5x} + q_{12}U_{xxx}U_x + q_{13}(U_{xx})^2 + q_{14}(U_x)^3, \\
U_{t_3} = {} & \varphi_3 + f_3 = - 6U^2U_x +  q_1 U^4 + q_2 U U_{xx} +
q_3 U_{xxx} + q_4 (U_x)^2, \\
U_{t_1} = {}& \varphi_1 + f_1 =  U_x.
\end{align*}
By solving the determining system of algebraic equations
$(U_{t_i})_{t_j} = (U_{t_j})_{t_i}$ upon the undetermined coefficients~$q_\alpha$
and then taking its nontrivial solution (if any), we obtain a new, dispersionful 
hierarchy (which is symmetry integrable by construction).
Specifically, for~\eqref{kdv} and its adjoint~\eqref{adjkdv} the solution is
\begin{align*}
\vdots&{}\\
U_{t_5} = {} & \tfrac{1}{30}\left(
  U_{4x} + 6U^5 + 10U^2U_{xx} + 10U(U_x)^2   \right)_x \\
U_{t_3} = {} & - U_{xxx} - 6U^2U_x,\\
U_{t_1} = {} & U_x,
\end{align*}
which is none other than the hierarchy of modified Korteweg\/--\/de Vries equation.
\end{example}

We now approach the main result of this report. Consider the second
term in one of the two towers of the Kaup\/--\/Boussinesq hierarchy,
\begin{subequations}\label{eqKBt2}
  \begin{align}
    u_{t_2} ={}& - v_{xxx} + 4(uv)_x,\\
    v_{t_2} ={}& - u_x + 4vv_x.
  \end{align}
\end{subequations}
Its Gardner deformation is known from~\cite{JMP2010}: let us recall that
the Miura contraction is 
\begin{subequations}\label{BLimMiura}
\begin{align}
u &= \tu + \veps\bigl(\tu_{x} - 2\tv\tv_{x}\bigr)
  + \veps^2\bigl(4\tu\tv^2 -\tu^2  - \tv_{x}^2\bigr)
+ 4\veps^3 \tu\tv\tv_{x}
 -4\veps^4  \tu^2\tv^2,\label{BLimMiurau12}\\
v &= \tv + \veps \tv_{x} - 2\veps^2\tu\tv,
   \label{BLimMiurau0}
\end{align}
\end{subequations}
and the parametric\/-\/extended equations are as follows,
\begin{subequations}\label{BLimBurgE}
\begin{align}
\tu_{t_2}&=\tv_{xxx}+4\bigl(\tu\tv\bigr)_x
 -2\veps\bigl(\tu_{x}\tv\bigr)_x
 -4\veps^2\bigl(\tu^2\tv\bigr)_x, \\
%\tfrac{\Id}{\Id x}\left[\frac14\tu_{0;xx}+\tu_{0}\tu_{12}\right] 
% + \epsilon \tfrac{\Id}{\Id x}\left[ -\frac12\tu_{0}\tu_{12;x} \right]
% + \epsilon^2 \tfrac{\Id}{\Id x}\left[ -\tu_{12}^2\tu_{0} \right]
\tv_{t_2} &= -\tu_{x}+4\tv\tv_{x}
 +2\veps\bigl(\tv\tv_{x}\bigr)_x
 -4\veps^2\bigl(\tu\tv^2\bigr)_x.
%\tfrac{\Id}{\Id x}\left[-\frac14\tu_{12} + \frac12\tu_{0}^2\right] 
% + \frac12\epsilon \tfrac{\Id}{\Id x} \left[\tu_{0}\tu_{0;x} \right] 
% + \epsilon^2 \tfrac{\Id}{\Id x}\left[ -\tu_{0}^2\tu_{12} \right]
\end{align}
\end{subequations}
By definition, the adjoint system is
\begin{equation}\label{eqKBt2adjoint}
\tu_{t_k} = (\tu^k\tv^{k-1})_x, \qquad \tv_{t_k} = (\tu^{k-1}\tv^k)_x.
\end{equation}

\begin{prop}\label{PropDeformKBous}
Let us require that the dispersionful extension of~\eqref{eqKBt2adjoint} 
itself is an infinite\/-\/dimensional integrable system and that it is 
scaling\/-\/invariant with respect to the weights
$[U]=[V]=\tfrac12$ and $[\Id/\Id t_k] = k$ for $k\in\mathbb{N}$.
Then there is a unique solution to the extension problem\textup{:} 
\begin{align*}
{}&\vdots &&\vdots {}\\
U_{t_3} &= (U_{xx} + 6UU_xV + 6U^3V^2)_x, & V_{t_3} &= (V_{xx} - 6UVV_x + 6U^2U^3)_x, \\
U_{t_2} &=  U_{xx} + 2 (U^2V)_x, & V_{t_2} &= - V_{xx} + 2(UV^2)_x, \\
U_{t_1} & = U_x, & V_{t_1} &= V_x.
\end{align*}
This is the Kaup\/--\/Newell hierarchy~\textup{\cite{KaupNewell}}.
\end{prop}

It would be quite logical to iterate the reasoning by first constructing a Gardner's
deformation --\,or several such deformations\,-- for the Kaup\/--\/Newell system, and then
by extending the available adjoint system(s).
However, this algorithmically simple problem appears unexpectedly complex 
as far as computations are concerned.
Specifically, by using the analytic software~\cite{SsTools} we obtain a 
`no-go` result: there is no Gardner's deformation for the
Kaup\/--\/Newell equation under the following set of assumptions:
\begin{itemize}
\item we supposed that the deformation $(\cE_\veps, \mathfrak{m}_{\veps})$ 
is polynomial in $\veps$ and differential polynomial in $\tilde{U}$ and~$\tilde{V}$;
\item we let such deformations be scaling homogeneous with respect to the weights
$[\tilde{V}] = [\tilde{U}] = \tfrac12$ and $[\veps] = - \tfrac12$;
\item the polynomial Ans\"atze for Gardner's deformations were bounded by using
$\deg_{\veps} (\mathfrak{m}_{\veps}) \leqslant 5$ and 
$\deg_{\veps} (\cE_{\veps}) \leqslant 10$
(here we note that $\max(\deg_{\veps} \mathfrak{m}_{\veps} )=
2\times\max(\deg_{\veps}(\cE_{\veps}))$ for the Kaup\/--\/Newell system).
\end{itemize}
Let as also note that the extended equation~$\cE_\veps$ for the Kaup\/--\/Newell system
can depend on derivatives of $U$ and $V$ with respect to $x$ of orders up to but not 
exceeding \emph{two}.\footnote{The proof is as follows: 
Consider the determining equation $(\mathfrak{m}_{\veps})_t = f(\mathfrak{m}_{\veps})$
for Gardner's deformation and calculate the differential orders of both sides;
by the chain rule, this yields that $\ord_x(\mathfrak{m}_{\veps}) + \ord_x(f_{\veps}) =
\ord_x(f)+ \ord_x(\mathfrak{m}_{\veps})$, which implies a rough estimate
$\ord_x(f_{\veps})=\ord_x(f)$.}
We expect that the Kaup\/--\/Newell system can be Gardner deformed strictly outside the 
class of differential polynomials (but can not be deformed within such class of functions).
  %exist only in the class of nonpolynomial Gardner's deformation.

It therefore remains an open problem to find Gardner's deformation(s) for 
the Kaup\/--\/Newell system and extend the arising adjoint equation(s)
so that new, higher\/-\/order completely integrable hierarchies are attained.

\medskip
%\subsection*{Conclusion}
We conclude that Gardner's deformations of infinite\/-\/dimensional completely integrable
systems can be effectively used not only through their `$\mathfrak{m}_\veps$-parts,' which 
encode the recurrence relations between conserved densities, but --\,viewed via their 
`$\cE_\veps$-\/parts'\,-- as a source of new completely integrable systems, 
or draft approximations to larger systems for which the integrability is retained.

The reproduction process is self\/-\/starting. Moreover, whenever there is
a Gardner deformation for the new hierarchy, one could attempt another iteration.
This scheme yields the oriented graph whose vertices are integrable systems and whose
edges associate new such systems to the ones at their starting points. We emphasize 
that the degree of a vertex can be greater than two,
meaning that a given system admits several deformations
(cf.~\cite{TMF2007,KiselevKarasu2006,JMP2010}), and that,
in principle, multiple edges may occur. A study of topology of such graph and
its correlation with the structure of moduli spaces for higher perturbations of 
low\/-\/order models is a challenging open problem.

%\subsection*{Acknowledgements}
\ack
The first author thanks the Organizing committee of international
workshop `Physics and Mathematics of nonlinear phenomena' (June
22\,--\,29, 2013; Gallipoli, Italy)  for stimulating discussions and
partial financial support. The research of A.V.K.\ was supported in
part by JBI~RUG project~103511 
(Groningen); A.O.K.\ was supported by ISPU scholarship for young scientists.
A~part of this research was done while the first author was visiting at
the $\smash{\text{IH\'ES}}$ (Bures\/-\/sur\/-\/Yvette);
the financial support and hospitality of this institution are gratefully
acknowledged.

\end{document}